# Privacy-preserving IoT Data Sharing Scheme

Ali Abdullah S. AlQahtani
Computer Systems Technology
North Carolina A&T State University
Greensboro, North Carolina
alqahtani.aasa@gmail.com

Hosam Alamleh
Computer Science
University of North Carolina Wilmington
Wilmington, North Carolina
hosam.amleh@gmail.com

Reem Alrawili
Applied Science and Technology
North Carolina A&T State University
Greensboro, North Carolina, USA
rfalrawili@aggies.ncat.edu

*Abstract*—Internet of Things (IoT) empowers us to connect physical objects ranging from smart buildings to portable intelligent devices, such as IoT wearable devices. The principal advantages of IoT wearable devices are allowing us remotely to collect data and also giving us the capability to control and monitor objects. These features can be utilized in various ways, such as establishing a connection, locating devices, authenticating users, protect users' privacy, sharing data and which is central to this proposed paper. Data sharing is receiving a widespread application in our daily lives since it facilitates cooperation between two ends, e.g., User-to-User, User-to-Devices, Devices-to-Devices, etc, and provides services to both ends or one of them. Data sharing can be granted using different factors, one of which is something in a user's/an IoT device's environment which is in this paper broadcast signals. Using broadcast signals to measure Received Signal Strength Indicator (RSSI) values and Machine Learning (ML) models, this paper implements an IoT data sharing scheme based on something that is in a user's/an IoT device's environment. The proposed scheme is experimentally tested using different ML models and shows 97.78% as its highest accuracy.

## I. INTRODUCTION

TODAY, IoT is emerging and is being adopted quickly. Such growth can is evident in numbers. Approximately, fifteen billion IoT devices worldwide were in use in 2015 [1] The number has doubled in 2020 (around 31 billion) and is expected to hit 60 billion by 2024 [2]. As a result, today IoT is considered as one of the most significant technologies in many sectors that may change the world [3] and has been a topic that interest many researchers. for example, Users' privacy and access control researchers have been studying IoT technologies and developing new products aiming to make the users' lives more convenient by utilizing smart IoT devices [4] that automate many of user's daily tasks and saves times and effort.

In order for smart devices in IoT to provide user-convenient services, it consumes a large amount of users' data. The amount of data that IoT devices consume and generate is staggering. According to a Federal Trade Commission report entitled "Internet of Things: Privacy & Security in a Connected World" , fewer than 10,000 households can generate 150 million discrete data points everyday. [5] This makes IoT a valuable target for hackers or even vendors. With that level of access to a user's home, much private and confidential information can be revealed or disclosed or revealed at different points in the systems. Therefore, data must be encrypted at all items until it is used. Furthermore, proper access control techniques must be in place. This paper tackles this issue by proposing a privacy preserving data sharing scheme.

This paper utilizes Bluetooth Low Energy(BLE) signal lev-els to achieve secure and privacy-preserving data sharing either between two users or between a user and an IoT device. In this paper, ML models are used to determine whether the two users or a user and an IoT device are proximate enough to share the data by utilizing their BLE signals levels. The proposed scheme is secure and limits data sharing to when the users or devices are proximate. Adding an extra layer of security and protecting the user's data privacy. The proposed scheme does not require any additional hardware as it uses a BLE-enabled user's device (whether wearable or smartphones).

This paper is organized as follows; Section II introduces the related works. Section III describes in details the proposed scheme. The experiment of the proposed scheme and its results are presented in Section IV. The proposed scheme is discussed in Section V. Security analysis of the proposed scheme is presented in Section VI Lastly, Section VII illustrates the conclusion for the proposed method.

## II. RELATED WORK

In this section, we review the related work, where discussed them in three different subsections. First, we discussed sharing data in IoT environments, then followed by the related works on end's privacy in IoT environments. Finally, we discussed usage of RSSI in determine objects' location.

### A. IoT Data Sharing

Two mechanisms were proposed for privacy-preserving data sharing in cloud-assisted IoT, which are encrypted data to a recipient [6], [7]. Moreover, the two papers introduced to the protection of confidentiality of data outsourced from IoT devices to other ends depending on cryptographic mechanisms.

In other words, collected and encrypted data can be accessed and read only by authorized requests. Blockchain technology can aid IoT techniques regarding privacy. Utilizing blockchain, IoT privacy-preserving and private data-sharing schemes were proposed [8], [9]. In the two discussed papers, secure data sharing can be accomplished automatically according to the predefined access permissions of data's owner through the smart contracts of blockchain.

### B. End's Privacy in IoT

A paper presents a privacy preferences expression framework for BLE-based applications named PrivacyBat [10]. The framework describes specifications for users to acquire arrangements on privacy rules with nearby BLE devices. Also, this framework delivers policies for a device to process user requests according to the arrangement. In another paper, the authors utilized a Voronoi diagram constructed by the Delaunay method to divide the road network space and decide the Voronoi grid region where the edge nodes are located [11]. Furthermore, they used a random disruption mechanism to fulfill the local differential privacy. Another system was proposed to protected personal location privacy in IOT environment from unauthorized access [12]. The system verifies the user's identity via the user authentication step and obtains. Then it can ensure that the user's location can be accessed only by an authorized user and only when required. Numerous of systems have addressed the seem concern and shown an acceptable results using different mechanisms [13]–[17].

### C. Usage of RSSI

As any signal travels form transmitter to receiver it decays which is represented by the RSSI and can be measured at the receiver end. Utilizing RSSI, we can estimate the distance between the transmitter and the receiver. RSSI has been utilized in privacy and access control fields more specifically in user authentication filed to prove users' identity [18]–[23]. In the mentioned papers, in order to gain access to a resource, a user must be within range of access points in his/her area, which can verified using RSSI values and estimate the distance between the user and the access points. Not only that but also, it shows a high accuracy in in the performance [24]–[26]. Moreover, RSSI has been used in determining objects' location to implement monitoring system [27]–[29].

## III. THE PROPOSED SCHEME

In this section, we will present the proposed scheme in two subsections: A. Registration; B. Accessing Data. In general, the proposed scheme validates the identity of each user in an IoT environment before one of them gain access to the other's private information.

### A. Registration Phase

In this phase, each user must be registered in the proposed scheme with a wearable IoT device, e.g., smartwatch, that has the ability to broadcast signals and measure RSSI values. Then, the scheme records the users' profile information along with more than one of the wearable IoT device's unique identifiers, e.g., UUID, IMEI, Device ID, etc... Based on the unique identifiers the proposed scheme will generate a unique signature for each device to be used in the communicating phase as an ID and identify themselves.

Based on their role, they will be classified into two groups, Group One and Group Two. Group One is the user/ device whom data is needed to be read, e.g., a student, a patient, or an IoT device. Group Two is the user who wants to read Group One user's data, e.g., a professor, a doctor or, an IoT user.

### B. Access Data Phase

The proposed scheme exploits the proximity by utilizing sensed information from the received BLE signals. Moreover, the proposed system utilizes the former non-connectable advertising mode that which means the devices don't have to be contacted with each other in the Access Data Phase. In the advertising mode, each BIE-enabled device broadcast their unique signature to identify themselves to each other.

A user from Group One must give a permission to a user from Group Two before he/she can access and read his/her data and this can be implemented following the steps below , which shows two use scenarios: (a) user B wants to access data to be shared by user A see Figure 1a, and (b) where user B wants to access data from an IoT device see figure 1b

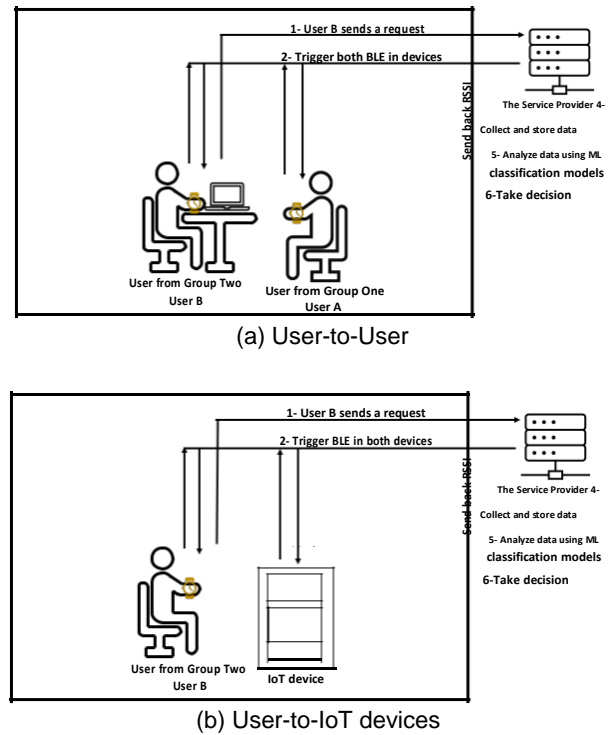

(a) User-to-User

(b) User-to-IoT devices

Fig. 1: Overview

1) A user from group two sends a request to the service provider asking for a permission to to access a user's data from group one.

2) The service provider triggers both BLE-enabled devices to communicate with each other; they identify them self using their unique signature.

3) They will measure the BLE RSSI values of each one's signals then send it to the service provider along with the timestamp.

4) The service provider collects the received data and stores them in a relational database at its end.

5) The the received data are then analyzed using ML classification models to determine their proximity

6) A decision is made to grant or deny permission for a user from Group Two.

## IV. Experiment

### A. Data collection

We used open-source available dataset [30]. The Data was collected in uncontrolled indoor environmental settings using two smartwatches that are embedded with BLE technology. Illustrative, An application was developed then installed in the two smartwatches to facilities sensing signals and measure RSSI values.

In the collection phase, Two participants were asked to;
1) Stand at a certain distance from each other; from 0.5m to 5m.
2) Wear the smartwatch on different hands and another in the same hand, which resulted to have four different settings, i.e., left to right (LR)), right to left (RL), left to left (LL), and right to right (RR).

The four settings were categorized into two categories, direct and crosswise. The direct is when the smartwatches are in a direct view; the crosswise is when the smartwatches are in a crosswise view. Figure 2 Visualizes the four settings and the two categorizations.

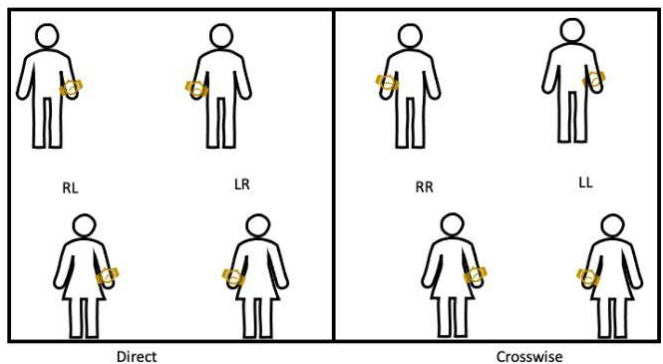

Fig. 2: The four settings visualization

37,644 data samples were collected from all the four different settings, see Table I. The collected data from RR and LL was combined as one dataset, i.e., crosswise dataset, and the collected data from RL and LR was combined as another dataset, i.e., direct dataset.

TABLE I: Total samples from each setting

| Settings | Number of Samples |
|---|---|
| Right to Right (RR) | 8168 |
| Left to Left (LL) | 7874 |
| Right to Left (RL) | 13117 |
| Left to Right (LR) | 8485 |

### B. Execution

Supervised learning was used to determine the proximity. We ran the two datasets, i.e., crosswise and direct datasets, through three different classifiers, i.e., Logistic Regression (LR), k-Nearest Neighbor (k-NN), Naive Bayes (NB), and Single Rule Indication (SRI). These three classifiers were selected since they provided the best results for the proposed scheme. In the execution phase, the data was split into 80% training and 20% testing; moreover, we calculated the system's accuracy, sensitivity, specificity, and precision.

Accuracy tells how many correct predictions to the total predictions, which helps to determine the best model for the proposed scheme. Sensitivity shows a scheme's ability to grant access to valid requests correctly. Conversely, specificity indicates a scheme's ability to correctly deny access to invalid requests. Finally, precision shows the number of legitimate accesses out of all accesses (legitimate and illegitimate).

We used Equations 1, 2, 3, and 4 to calculate the results in Tables II and III.

$$\text{Accuracy} = \frac{True\ Positive + True\ Negative}{Total\ number\ of\ a\ dataset} \times 100 \quad (1)$$

$$\text{Sensitivity} = \frac{True\ Positive}{True\ Positive + False\ Negative} \times 100 \quad (2)$$

$$\text{Specificity} = \frac{True\ Negative}{True\ Negative + False\ Positive} \times 100 \quad (3)$$

$$\text{Precision} = \frac{True\ Positive}{True\ Positive + False\ Positive} \times 100 \quad (4)$$

TABLE II: Crosswise Results

| ML Model | LR | KNN | NB |
|---|---|---|---|
| Accuracy | 97.29% | 94.86% | 85.50% |
| Sensitivity | 97.79% | 96.64% | 88.97% |
| Specificity | 96.66% | 92.62% | 81.14% |
| Precision | 97.35% | 94.27% | 85.56% |

TABLE III: Direct Results

| ML Model | LR | KNN | NB |
|---|---|---|---|
| Accuracy | 97.78% | 95.41% | 90.59% |
| Sensitivity | 98.35% | 97.49% | 92.44% |
| Specificity | 97.03% | 92.66% | 88.14% |
| Precision | 97.77% | 94.62% | 91.17% |

## V. System Analysis

### A. Seamless Access control

The proposed scheme implements seamless access control, which can be done without the user's interaction. For example, data can be shared as soon as the user is close to the IoT device or when a user wants to share data with another user. This is important in the IoT context as such systems must provide a high level of user convince, otherwise, they won't be used. The proposed scheme increases users' privacy without affecting users'experience.

### B. Usability

Due to the proposed scheme using the existing BLE-enabled devices, (i.e., BLE is already in smartphones, wearable, IoT devices), no new hardware is required to be installed. Moreover, the proposed scheme is considered to be readily applicable in a scalable manner. Today , BLE is inexpensive (i.e, a BLE chip costs between 1-2 dollars) and consumes little energy therefore it is implemented in a wide range of devices and equipment, which makes the proposed scheme applicable to a wide range of scenarios.

## VI. Security Analysis

### A. Environment Simulation Attack

A chance of simulating the IoT Environment is possible, where an attacker scans the IoT environment (i.e., BLE advertisements in the area) and then replicates it elsewhere, either using relay or replay attacks or by cloning a device that has the same broadcast frame. However, in real-world scenarios, an authentication factor is already in place, which authenticates users to use the services (i.e, an account with a username and password). Therefore, this attack scenario can be mitigated by requiring users to be signed in to the service.

### B. Accidental Unauthorized Access

In this scenario, where the user is proximate to another device or to another user, and an attacker uses this proximity to get access to the data without the user's knowledge. This attack scenario can be mitigated by requiring the user to perform an action to access the data (e.g, press a button on their smartphone or perform a gesture on a wearable). This could affect user convenience. However, such requirements can be determined by user preference and/or the data sensitivity (i.e, for sensitive data performing an action is required and not for insensitive data).

## VII. Conclusion

IoT empowers users to connect physical objects ranging from smart buildings to portable intelligent devices, such as IoT wearable devices. The principal advantages of IoT wearable devices are allowing us remotely to collect data and also giving us the capability to control and monitor objects. These features can be utilized in various ways, such as establishing a connection, locating devices, authenticat-ing users, protect users' privacy, sharing data and which is central to this proposed paper. Data sharing is receiving a widespread application in our daily lives since it facilitates cooperation between two ends, e.g., User-to-User, User-to-Devices, Devices- to-Devices, etc, and provides services to both ends or one of them. Data sharing can be granted using different factors, one of which is something in a user's/an IoT device's environment which is in this paper broadcast signals. In this paper, ML models and BLE signal levels are utilized to determine whether the two users, or the user and the IoT device are proximate enough to share the data. We experimentally tested the proposed scheme using different ML models that have provided a satisfactory results, which shows 97.78% as its highest accuracy. Moreover, different aspects from the proposed scheme was discussed and analysed.